\documentstyle[aps]{revtex}
 \def\ket{\rangle} 
\def\<{\langle}
\def\>{\rangle}

\begin{document}
\title{General Schemes for multi-particle d-dimensional Cat-like State Teleportation}
\author{
Bei Zeng$^{1,2}$, Xiao-Shu Liu$^{3,1,2}$, Yan-Song Li$^{1,2}$, Gui Lu Long$^{1,2}$}
\address{
$^1$ Department of Physics, Tsinghua University,  Beijing 100084, P R China\\
$^2$ Key Laboratory For Quantum Information and Measurements, Beijing 100084, P R China\\
$^3$ Department of Physics, Shandong Normal University, Jinan 250014, P R China}
\maketitle    
\begin{abstract}
Two kinds of $M$-particle d-dimensional cat-like state teleportation protocols are 
present. In the first protocol, the teleportation is achieved by d-dimensional Bell-basis 
measurements, while in the second protocol it is realized by d-dimensional GHZ-basis 
measurement. It is also shown that the second protocol has a simple mathematical 
formulation that is identical to Bennett et al's original teleportation protocol for an 
unknown state of a single particle.  
\end{abstract}

Since first suggested by Bennett et al\cite{r1}, teleportation of quantum state has 
attracted the attention of both theorists and experimentalists. It is one of the most 
important aspect of quantum information.  In their original work, Bennett et al gave a 
protocol to teleport an unknown quantum state for qubit system using EPR pairs. Besides 
the qubit case, they also generalized the scheme to the case where the single particle has 
d-dimensional Hilbert space using a pair of d-state particles in a maximally entangled 
state $\sum\limits_i|i\>\otimes|i\>/\sqrt d$, where i=0, 1, 2,$\ldots$, d-1. This protocol
 is studied in detail recently\cite{r2} using positive operator valued measurement. 
Meanwhile, Bennett et al also considered the teleportation 
 of $M$-particle state. Their scheme is actually to teleport particles one-bye-one, and it 
requires $2M$ particles to teleport $M$ particle state. Multipartite  cat-like state 
teleportation raises high attention recently\cite{r3,r4,r5,r6,r7}, for it is found that 
for qubit system they can be teleported using fewer particles than Bennett et al's 
protocol, i.e. $M$-particle cat-like state need only $M+1$ particles to teleport compared 
to Bennett's $2M$ particles. 
The core problem to construct these protocols is to find a prorate measurement 
basis Alice should choose. 

In this paper, we present two protocols to teleport $M$-particulate state in d-dimension. 
One protocol is the generalization of the standard protocol associated with Bell-basis 
measurement\cite{r3}, and the other protocol is associated with the GHZ-basis 
measurement\cite{r4}. We will begin with simple examples and then give the explicit 
expression for the general case. In particular, the $M$-particle case will be considered 
and the relation between the two protocols and those with the original protocol of Bennett 
et al. will be analyzed.

{\bf First we consider the teleportation of cat-like state of 2 particles in 3-dimension 
using  ``Bell-basis measurement'' protocol}. 
Alice has a 3-dimensional cat-like state $|\Psi_{cat}\>$ which she wants to 
teleport to Bob:
\begin{equation}
|\Psi_{cat}\>=\alpha|0\>_1|0\>_2+\beta|1\>_1|1\>_2+\gamma|2\>_1|2\>_2
\end{equation}
As is common to cat-like teleportation study, the basis $|00\>$, $|11\>$ and $|22\>$ are 
known, but the coefficients $\alpha$, $\beta$ and $\gamma$ are unknown.

We use  maximally entangled state of three particle $|\Psi_M\>$ to complete our task:
\begin{equation}
|\Psi_M\>=(|0\>_3|0\>_4|0\>_5+|1\>_3|1\>_4|1\>_5|+|2\>_3|2\>_4|2\>_5)/\sqrt3.
\end{equation}
We gave particle 3 to Alice and particles 4 and 5 to Bob. Thus the state of the whole 
system is
\begin{equation}
|\Psi\>=\frac1{\sqrt3}(\alpha|0\>_1|0\>_2+\beta|1\>_1|1\>_2+\gamma|2\>_1|2\>_2)\otimes
(|0\>_3|0\>_4|0\>_5+|1\>_3|1\>_4|1\>_5|+|2\>_3|2\>_4|2\>_5).
\end{equation}
``Bell-basis measurement'' requires the  collapse of the second and third particle to the 
state\cite{r1}:
\begin{equation}
|\Psi_{nm}\>=\sum\limits_je^{2\pi ijn/3}|j\>\otimes|j+m\,mod\,3\>/\sqrt3
\end{equation}
where n, m, j=0, 1, 2. More explicitly:
\begin{equation}\begin{array}{l}
|\Psi_{00}\>=(|00\>+|11\>+|22\>)/\sqrt3,\\
|\Psi_{10}\>=(|00\>+e^{2\pi i/3}|11\>+e^{4\pi i/3}|22\>)/\sqrt3,\\
|\Psi_{20}\>=(|00\>+e^{4\pi i/3}|11\>+e^{2\pi i/3}|22\>)/\sqrt3,\\

|\Psi_{01}\>=(|01\>+|12\>+|20\>)/\sqrt3,\\
|\Psi_{11}\>=(|01\>+e^{2\pi i/3}|12\>+e^{4\pi i/3}|20\>)/\sqrt3,\\
|\Psi_{21}\>=(|01\>+e^{4\pi i/3}|12\>+e^{2\pi i/3}|20\>)/\sqrt3,\\

|\Psi_{02}\>=(|02\>+|10\>+|21\>)/\sqrt3,\\
|\Psi_{12}\>=(|02\>+e^{2\pi i/3}|10\>+e^{4\pi i/3}|21\>)/\sqrt3,\\
|\Psi_{22}\>=(|02\>+e^{4\pi i/3}|10\>+e^{2\pi i/3}|21\>)/\sqrt3.\\
\end{array}\end{equation}
However, to complete our task, a combined measurement of Alice's three particles is 
needed. The combined measurement includes a Bell-basis measurement for particles 2 and 3, 
and a separate measurement on particle 1 in the  $\{|\pi_0\>,|\pi_1\>,|\pi_2\>\}$ basis 
defined by:
\begin{equation}
\left[\begin{array}{c}|\pi_0\>\\|\pi_1\>\\|\pi_2\>\end{array}\right]=\frac1{\sqrt3}
\left[\begin{array}{ccc}1&1&1\\1&e^{2\pi i/3}&e^{4\pi i/3}\\1&e^{4\pi i/3}&e^{2\pi 
i/3}\end{array}\right]\left[\begin{array}{c}|0\>\\|1\>\\|2\>\end{array}\right].
\end{equation}

The state of the whole system can be rewritten in the form:
\begin{equation}
|\Psi\>=\sum\limits_{n,m,k}|\pi_k\>_1\otimes|\Psi_{nm}\>_{23}
\otimes|\varphi^k_{nm}\>_{45}.
\end{equation}
Thus, after the combined measurement if Alice obtains the result that particles 2 and 3 
are in state $|\Psi_{nm}\ket$, and particle 1 is in state $|\pi_k\ket$, the unitary 
transformation $U^k_{nm}$
\begin{equation}
U^k_{nm}=\frac13\sum\limits_je^{2\pi 
ij(n+k)/3}|j\>\otimes|j\>\<(j+m)\,mod\,3|\otimes\<(j+m)\,mod\,3|
\end{equation}
will transform $|\varphi^k_{nm}\>_{45}$ to 
$|\Psi_{cat}=\alpha|0\>_4|0\>_5+\beta|1\>_4|1\>_5+\gamma|2\>_4|2\>_5,$
which completes the teleportation. 

{\bf Teleportation of cat-like state of 2 particles in d-dimension using Bell-basis 
protocols} is straightforward. To teleport the d-dimensional cat-like state
\begin{equation}
|\Psi_{cat}\>=\sum\limits_l\alpha_l|l\>_1l\>_2\hspace{1cm}l=0,\,1,\,\ldots,\,d-1,
\end{equation}
we only need to collapse the particles 2, 3 to the basis
\begin{equation}
|\Psi_{nm}\>=\sum\limits_je^{2\pi ijn/d}|j\>\otimes|j+m\,mod\,d\>/\sqrt d
\end{equation}
and simultaneously collapse particle 1 to the basis $\{|\pi_\alpha\>\}$
\begin{equation}
|\pi_\alpha\>=\frac1{\sqrt d}\sum\limits_{\beta=0}^{d-1}e^{2\pi 
i\alpha\beta/d}|\beta\>\hspace{1cm}\alpha=0,\,1,\,\ldots,\,d-1.
\end{equation}
Using these basis, the state of the whole system can be rewritten in the form:
\begin{equation}
|\Psi\>=\sum\limits_{n,m,k}|\pi_k\>_1\otimes|\Psi_{nm}\>_{23}
\otimes|\varphi^k_{nm}\>_{45}.
\end{equation}
Thus, after the combined measurement of particles 1, 2, and 3 by Alice, if the results of 
the measurement is particles 2 and 3 are in state $|\Psi_{nm}\ket$ and particle 1 in state 
$|\pi_k\ket$, the unitary transformation $U^k_{nm}$
\begin{equation}
U^k_{nm}=\frac1d\sum\limits_je^{2\pi 
ij(n+k)/d}|j\>\otimes|j\>\<(j+m)\,mod\,d|\otimes\<(j+m)\,mod\,d|
\end{equation}
will transform $|\varphi^k_{nm}\>_{45}$ to $|\Psi_{cat}\>_{45}$.

{\bf Furthermore, the above results can be generalized in the many particle case}. For 
$M$-particle cat-like state with d-dimension
\begin{equation}
|\Psi_{cat}\>=\sum\limits_{l=0}^{d-1}\alpha_l\prod\limits_{i=1}^M|l\>_i.
\end{equation}
It is not difficult to complete the teleportation task with other M+1 particles in the 
maximally entangled state as
\begin{equation}
|\Psi_{M+1}\>=\frac1{\sqrt d}\sum\limits_{i=0}^{d-1}|\underbrace{ii\cdots 
i}_{M+1}\>_{M+1,\ldots,2M+1}.
\end{equation}
Then, as a whole, the state of the system can be written as:
\begin{equation}
|\Psi\>=|\Psi_{cat}\>\otimes|\Psi_{M+1}\>=\sum\limits_l\alpha_l|\underbrace{ll\cdots 
l}_M\>\otimes\sum\limits_i|\underbrace{ii\cdots i}_{M+1}\>/\sqrt 
d=\sum\limits_{k,n,m,\alpha_i}|\pi_{\alpha_1\alpha_2\cdots\alpha_{M-1}}\>\otimes|\Psi_{nm}
\>\otimes|\varphi_{nm}^{\alpha_1\alpha_2\cdots\alpha_{M-1}}\>,
\end{equation}
where 
$|\pi_{\alpha_1\alpha_2\cdots\alpha_{M-1}}\>=\prod\limits_{i=1}^{M-1}|\pi_{\alpha_i}\>_i$ 
and $\alpha_i=0,\,1,\,\ldots,\,d-1$.
 We only need to make a measurement to collapse the M, M+1 particles into the 
 basis $\{|\Phi_{mn}\>\}$ and measure particles 1 to M on the 
 basis $\{|\pi_{\alpha_1\alpha_2\cdots\alpha_{M-1}}\>\}$ simultaneously. After the results 
of the measurement is known, Bob applies the  
 unitary transformation $U_{nm}^{\alpha_1\alpha_2\cdots\alpha_{M-1}}$ defined by
\begin{equation}
U_{nm}^{\alpha_1\alpha_2\cdots\alpha_{M-1}}=\sum\limits_je^{2\pi 
ij(\sum\limits_{i=1}^{n-1}\alpha_i)/d}|j\>\otimes|j\>\<(j+m)\,mod\,d|\otimes\<(j+m)\,mod\,
d|/d
\end{equation}
to his $M$-particle which is in state 
$|\varphi^{\alpha_1\alpha_2\cdots\alpha_{M-1}}_{nm}\>$, and this  will transform his state 
to $|\Psi_{cat}\>$. This complements the teleportation task.

{\bf Now we turn to the ``GHZ-basis measurement'' protocol}. We first discuss the case of 
2 particles in 3-dimension, and then we generalize the 2 partcle case into d-dimensions. 
Finally we give the results for the general case of $M$-particles in d-dimensions.

{\bf First we consider the 2 particle in 3-dimension case in detail}. Alice possesses a 
3-dimensional cat-like 
state $|\Psi_{cat}\>$ which she wants to teleport to Bob:
\begin{equation}
|\Psi_{cat}\>=\alpha|0\>|0\>+\beta|1\>|1\>+\gamma|2\>|2\>.
\end{equation}
We use three particles in the  maximally entangled state $|\Psi_M\>$ to complete our task:
\begin{equation}
|\Psi_M\>=(|0\>|0\>|0\>+|1\>|1\>|1\>|+|2\>|2\>|2\>)/\sqrt3
\end{equation}
Then we choose the collapse basis for particle 1, 2, 3 as
\begin{equation}
|\Psi_{nm}^k\>=\sum\limits_je^{2\pi 
ij(n+k)/3}|j\>\otimes|(j+n)\,mod\,3\>\otimes|(j+m)\,mod\,3\>/\sqrt3.
\end{equation}
In this basis,  the state of the whole system can be written in the form:
\begin{equation}
[\alpha|0\>|0\>+\beta|1\>|1\>+\gamma|2\>|2\>]\otimes[(|0\>|0\>|0\>+|1\>|1\>|1\>|+|2\>|2\>|
2\>)/\sqrt3]=\sum\limits_{k,m}|\Psi_{0m}^k\>\otimes|\varphi_m^k\>,
\end{equation}
where
\begin{equation}\begin{array}{ll}
&[\alpha|0\>|0\>+\beta|1\>|1\>+\gamma|2\>|2\>]\otimes[(|0\>|0\>|0\>+|1\>|1\>|1\>|+|2\>|2\>
|2\>)/\sqrt3]\\
=&|\Psi_{00}^0\>\otimes[\alpha|00\>+\beta|11\>+\gamma|22\>]\\
+&|\Psi_{00}^1\>\otimes[\alpha|00\>+e^{4\pi i/3}\beta|11\>+e^{2\pi i/3}\gamma|22\>]\\
+&|\Psi_{00}^2\>\otimes[\alpha|00\>+e^{2\pi i/3}\beta|11\>+e^{4\pi i/3}\gamma|22\>]\\
+&|\Psi_{01}^0\>\otimes[\alpha|00\>+\beta|11\>+\gamma|22\>]\\
+&|\Psi_{01}^1\>\otimes[\alpha|00\>+e^{4\pi i/3}\beta|11\>+e^{2\pi i/3}\gamma|22\>]\\
+&|\Psi_{01}^2\>\otimes[\alpha|00\>+e^{2\pi i/3}\beta|11\>+e^{4\pi i/3}\gamma|22\>]\\
+&|\Psi_{02}^0\>\otimes[\alpha|00\>+\beta|11\>+\gamma|22\>]\\
+&|\Psi_{02}^1\>\otimes[\alpha|00\>+e^{4\pi i/3}\beta|11\>+e^{2\pi i/3}\gamma|22\>]\\
+&|\Psi_{02}^2\>\otimes[\alpha|00\>+e^{2\pi i/3}\beta|11\>+e^{4\pi i/3}\gamma|22\>]\\
\end{array},
\end{equation}
and $|\varphi_m^k\>$ is the state of Bob's particles. Thus, after the measurement made by 
Alice in the above basis, the unitary transformation $U_m^k$
\begin{equation}
U^k_m=\frac13\sum\limits_je^{2\pi 
ijk/3}|j\>\otimes|j\>\<(j+m)\,mod\,3|\otimes\<(j+m)\,mod\,3|
\end{equation}
will transform Bob's state $|\varphi_m^k\>$ to $|\Psi_{cat}\>$.

{\bf Generalization to d-dimensional case is also straightforward}. To teleport the 
d-dimensional cat-like state
\begin{equation}
\label{e1}
|\Psi_{cat}\>=\sum\limits_l\alpha_l|l\>l\>\hspace{1cm}l=0,\,1,\,\ldots,\,d-1,
\end{equation}
we need to collapse the 1, 2, 3 particles to the basis
\begin{equation}
\label{e2}
|\Psi_{nm}^k\>=\sum\limits_je^{2\pi 
ij(n+k)/d}|j\>\otimes|(j+n)\,mod\,d\>\otimes|(j+m)\,mod\,d\>/\sqrt d.
\end{equation}
In this basis the  state of the whole system can be written in the form:
\begin{equation}
\label{e3}
|\Psi\>=\sum\limits_{k, m}|\Psi^k_{0m}\>\otimes|\varphi_m^k\>.
\end{equation}
Thus, after the GHZ-type of measurement by Alice which collapse Alice's partciles to 
states labeled by $n$, $m$ and $k$, the unitary transformation
\begin{equation}
\label{e4}
U_m^k=\sum\limits_je^{2\pi ijk/d}|j\>\otimes|j\>\<(j+m)\,mod\,d|\otimes\<(j+m)\,mod\,d|/d
\end{equation}
will transform Bob's partcle state $|\varphi_m^k\>$ to $|\Psi_{cat}\>$.

{\bf The generalization to the $M$-particle case is very easy.} To teleport an 
$M$-particle d-dimension ct-like state,
\begin{equation}
\label{e5}
|\Psi_{cat}\>=\sum\limits_la_l|l\>_1\otimes|\underbrace{ll\cdots 
l}_{M-1}\>_{2,3,\ldots,M},\hspace{1cm}l=0,\,1,\,\ldots,\,d-1,
\end{equation}
using $M+1$ particles in the maximally entangled state, we can make the following 
observations. 
Denoting $|l\>_{\overline2}=|\underbrace{ll\cdots l}_{M-1}\>_{2,3,\ldots,M}$, the cat-like 
state becomes
\begin{equation}
|\Psi_{cat}\>=\sum\limits_la_l|l\>_1\otimes|l\>_{\overline2}.
\end{equation}
By  by replacing $|l\>_2$ with $|l\>_{\overline2}$ and $|(j+n)\,mod\,2\>_2$ with 
$|(j+n)\,mod\,2\>_{\overline2}$ in (\ref{e1}-\ref{e4}), we obtain the basis to collapse 
for Alice and the corresponding transformation for Bob. In this respect we can say that 
the teleportation of M-particle cat-like state is equivalent to 2-particle cat-like state 
teleportation in the ``GHZ-basis measurement'' protocol.

Finally, it should be pointed out that {\bf all the ``GHZ-basis measurement'' protocols
can  be reduced to Bennett et al.'s original protocol for a single particle}
\cite{r1}. In fact, 
\begin{equation}
|\overline l\>=|\underbrace{ll\cdots l}_M\>_{1,2,\ldots,M}
\end{equation}
resembles a ''single'' particle state in Bennett et al's protocol, the $M+1$ particle 
state
\begin{equation}
|\overline\Psi_{nm}\>=\sum\limits_je^{2\pi ijn/d}|\overline 
j\>\otimes|{(j+m)\,mod\,d}\>/\sqrt d,
\end{equation}
resembles the ``2'' particle states in Bennett's protocol, and
\begin{equation}
\overline U_{nm}=\sum\limits_ke^{2\pi ikn/d}|\overline k\>\<\overline{(k+m)\,mod\,d}|.
\end{equation}
is the unitary transformation of Bob on his ''single'' particle. In this way, we see that 
the mathematical expressions are identical for $M$-particle cat-like teleportation and the 
single particle teleportation protocol.

Therefore, any cat-like teleportation problem can be solved when measuring in the basis 
constituting of
$\{|\overline\Psi_{nm}\>\}$ and its orthogonal complement followed by applying the unitary 
$\overline U_{nm}$ (It is noted that only the coefficients of 
$\{|\overline\Psi_{nm=0m}\>\}$ are nonzero). Thus, accurate teleportation of $M$-particle 
with d-state requires $2\log_2M$ bits in ``GHZ-basis measurement'' protocol. This is not 
surprising because the cat-like state that we want to teleport belongs to the 
d-dimensional subspace of the whole Hilbert space. Mathematically we can well treat the 
$M$-particle d-dimension unknown cat-like state 
teleportation problem as the teleportation of an unknown d-dimensional state of a single 
particle. We can understand the result of Werner\cite{r8} that the ``GHZ-basis 
measurement'' protocol is the optimal one. Because we know it costs the fewest classical 
communication among all the possible faithful protocols. 

Regarding implementation,  it is noted the ``GHZ-basis measurement'' protocol is harder to 
implement experimentally than the Bell-basis measurement protocol, for it needs to do 
$M$-particle collective measurement. While 
``Bell-basis measurement'' protocol is much easier to implement experimentally\cite{r9}, 
although it costs $(M+1)\log_2M$ classical bits. It is easy to construct other protocols 
with $k\log_2M$ classical bits $2\leq k \leq (M+1)$, when Alice chooses to do
$(M-k+3)$-particle collective measurement.  

This work is supported by the Major State Basic Research Developed Program 
Grant No. G200077400, the China National Natural Science Foundation Grant 
No. 60073009, the Fok Ying Tung Education Foundation, and the 
Excellent Young University Teachers' Fund of Education Ministry of China. 

\end{document}